\documentclass{ws-ijmpe}

\usepackage[normalem]{ulem} 
\usepackage[dvips]{color} 
\renewcommand\sout{\bgroup \color{red} \ULdepth=-.5ex \ULset}
\newcommand{\com}[1]{{\sf\color[rgb]{0,0,1}{#1}}}

\begin{document}

\markboth{C. M. Ko, Y. Oh, and J. Xu}{Medium effects on charged pion ratio in heavy ion collisions}

\catchline{}{}{}{}{}

\title{MEDIUM EFFECTS ON CHARGED PION RATIO IN HEAVY ION COLLISIONS}

\author{CHE MING KO}

\address{Cyclotron Institute and Department of Physics and Astronomy, Texas A\&M University, College Station, Texas 77843-3366\\
ko@comp.tamu.edu}

\author{YONGSEOK OH}

\address{Korea Institute of Science and Technology Information, Daejeon 305-806, Korea\\
yoh@kisti.re.kr}

\author{JUN XU}

\address{Cyclotron Institute, Texas A\&M University, College Station, Texas 77843-3366\\
xujun@comp.tamu.edu}

\maketitle

\begin{history}
\received{(received date)}
\revised{(revised date)}
\end{history}

\begin{abstract}
We have recently studied in the delta-resonance--nucleon-hole model
the dependence of the pion spectral function in hot dense asymmetric
nuclear matter on the charge of the pion due to the pion $p$-wave
interaction in nuclear medium.  In a thermal model, this
isospin-dependent  effect enhances the ratio of negatively charged
to positively charged pions in neutron-rich nuclear matter, and the
effect is comparable to that due to the uncertainties in the
theoretically predicted stiffness of nuclear symmetry energy at high
densities. This effect is, however, reversed if we also take into
account the $s$-wave interaction of  the pion in nuclear medium as
given by chiral perturbation theory, resulting instead in a slightly
reduced ratio of negatively charged to positively charged pions.
Relevance of our results to the determination of the nuclear
symmetry energy from the ratio of negatively to positively charged
pions produced in heavy ion collisions is discussed.
\end{abstract}

\section{Introduction}

The nuclear symmetry energy is the energy needed per nucleon to
convert all protons in a symmetric nuclear matter to neutrons.
Knowledge on the density dependence of nuclear symmetry energy is
important for understanding the dynamics of heavy ion collisions
induced by radioactive beams, the structure of exotic nuclei with
large neutron or proton excess, and many important issues in nuclear
astrophysics~\cite{LCK08,ireview98,ibook,baran05}. At normal nuclear
matter density, the nuclear symmetry energy has long been known to
have a value of about $30$~MeV from fitting the binding energies of
atomic nuclei with the liquid-drop mass formula. Somewhat stringent
constraints on the nuclear symmetry energy below the normal nuclear
density have also been obtained during past few years from studies
of the isospin diffusion~\cite{Tsa04,Liu07,Che05a,LiBA05c} and
isoscaling~\cite{She07} in heavy-ion reactions, the size of neutron
skins in heavy nuclei~\cite{Ste05b}, and the isotope dependence of
giant monopole resonances in even-A Sn isotopes~\cite{Gar07}.  For
nuclear symmetry energy at high densities, transport model studies
have shown that the ratio of negatively to positively charged pions
produced in heavy ion collisions with neutron-rich nuclei is
sensitive to its stiffness~\cite{bali02,Ferini:2006je}. Comparison
of this ratio from an isospin-dependent Boltzmann-Uehling-Ulenbeck
(IBUU) transport model based on the non-relativistic
momentum-dependent (MDI) nuclear effective
interactions~\cite{xiao09} with measured data from heavy ion
collisions by the FOPI Collaboration~\cite{FOPI} at GSI seems to
indicate  that the nuclear symmetry energy at high density might be
very soft.  Although this study does not include the relativistic
effects,  which may affect the charged pion ratio as shown in
Ref~\cite{Ferrini:2005jw}, it provides an important step in the
determination of the nuclear symmetry energy at high densities.

The transport model used in Ref.~\cite{xiao09} neglects, however,
medium effects on pions, although it includes those on nucleons and
produced $\Delta$ resonances through their isospin-dependent
mean-field potentials and scattering cross sections. It is
well-known that pions interact strongly in nuclear medium as a
result of their $p$-wave couplings to the
nucleon-particle--nucleon-hole and delta-particle--nucleon-hole
($\Delta$-hole) excitations, leading to the softening of their
dispersion relations or an increased strength of their spectral
functions at low
energies~\cite{weise75,friedmann81,oset82,xia94,hees05,korpa08}.
Including pion medium effects in the transport model has previously
been shown to enhance the production of low energy pions in high
energy heavy ion collisions, although it does not affect the total
pion yield~\cite{xiong93}. Since pions of different charges are
modified differently in asymmetric nuclear matter that
has unequal proton and neutron fractions~\cite{korpa99},
including such isospin-dependent medium effects is expected to
affect the ratio of negatively to positively charged pions produced
in heavy ion collisions.

\section{Pion $p$-wave interactions in nuclear medium}

Considering only the dominant $\Delta$-hole excitations as in
Ref.~\cite{ko89}, as the contribution from the nucleon particle-hole
excitations is known to be small, the self-energy of a pion of
isospin state $m_t$, energy $\omega$, and momentum $k$ in a hot
nuclear medium due to its $p$-wave interaction is given by
\begin{eqnarray}\label{pi}
\Pi_0^{m_t} &\approx& \frac{4}{3} \left(
\frac{f_\Delta^{}}{m_\pi^{}} \right)^2 k^2 F_\pi^2(k)
\sum_{m_\tau,m_T^{}} \left|\left\langle {\textstyle\frac{3}{2}} \,
m_T^{} | 1\, m_t\, {\textstyle\frac{1}{2}} \, m_\tau \right\rangle\right|^2\notag\\
&\times& \int \frac{d^3p}{(2\pi)^3}\frac{1}{e^{(m_N^{} + p^2/2m_N^{}
+U_N^{m_\tau}-\mu_B^{}-2m_\tau\mu_Q^{})/T}+1}
\left(\frac{1}{\omega-\omega_{m_T^{}}^{+}}+\frac{1}{-\omega-\omega_{m_T^{}}^{-}}\right),\notag\\
\end{eqnarray}
with $\omega_{m_T^{}}^{\pm} \approx m_\Delta^{} +U_\Delta^{m_T^{}} +
(\vec{k} \pm \vec{p})^2/2m_\Delta^{}-i\Gamma_\Delta^{m_T^{}}/2
-m_N^{}-U_N^{m_\tau} - p^2/2 m_N^{}$. In the above, $m_\pi \simeq
138$~MeV, $m_N^{} \simeq 939$~MeV, and $m_\Delta^{} \simeq 1232$~MeV
are the masses of pion, nucleon, and $\Delta$ resonance,
respectively; $f_\Delta^{} \simeq 3.5$ is the $\pi N\Delta$ coupling
constant and $F_\pi(k) = [1+0.6(k^2/m^2_\pi)]^{-1/2}$~\cite{art} is
the $\pi N\Delta$ form factor determined by fitting the decay width
$\Gamma_\Delta\simeq 118$~MeV of $\Delta$ resonance in free space.
The summation in Eq.~(\ref{pi}) is over the nucleon isospin state
$m_\tau$, and the $\Delta$ resonance isospin state $m_T^{}$; and the
factor $\langle {\textstyle\frac{3}{2}} \, m_T^{} | 1\,
m_t\,{\textstyle\frac{1}{2}} \, m_\tau \rangle$ is the
Clebsch-Gordan coefficient from the isospin coupling of pion with
nucleon and $\Delta$ resonance. The momentum integration is over
that of nucleons in the nuclear  matter given by a Fermi-Dirac
distribution with $\mu_B$ and $\mu_Q$ being,  respectively, the
baryon and charge chemical potentials determined by  charge and
baryon number conservations; $\rho_N^{m_\tau}$ and $U_N^{m_\tau}$
are, respectively, the density and mean-field potential of nucleons
of isospin state $m_\tau$ in asymmetric nuclear matter; and
$\Gamma_\Delta^{m_T^{}}$ and $U_\Delta^{m_T^{}}$ are, respectively,
the width and mean-field potential of $\Delta$ resonance of isospin
state $m_T^{}$.

For the nucleon mean-field potential $U_N^{m_\tau}$, we have used
the one obtained from the momentum-independent (MID)
interaction~\cite{LCK08}, i.e., $U_N^{m_\tau}(\rho_B^{},\delta_{\rm like}) =
\alpha(\rho_B^{}/\rho_0^{}) + \beta(\rho_B^{}/\rho_0^{})^\gamma +
U_{\text{asy}}^{m_\tau}(\rho_B^{} ,\delta_{\rm like})$, with
$U_{\text{asy}}^{m_\tau}(\rho_B^{} ,\delta_{\rm like})= -4\{ F(x)(\rho_B^{}/\rho_0^{}) + [18.6 -
F(x)] (\rho_B^{}/\rho_0^{})^{G(x)}\} m_\tau \delta_{\rm like} + [18.6 -
F(x)][G(x) - 1] (\rho_B^{}/\rho_0^{})^{G(x)} {\delta_{\rm
like}}^{2}$ being the nucleon symmetry potential. The parameters
$\alpha=-293.4$~MeV, $\beta=240.1$~MeV, and $\gamma=1.216$ are
chosen to give a compressibility of $212$~MeV and a binding energy
per nucleon of $-16$~MeV for symmetric nuclear matter at the
saturation or normal nuclear density $\rho_0^{} = 0.16~{\rm
fm}^{-3}$. The nucleon symmetry potential
$U_{\rm asy}^{m_\tau}(\rho_B^{} ,\delta_{\rm like})$
depends on the baryon density $\rho_B^{} = \rho_n^{} + \rho_p^{} +
\rho_{\Delta^-}^{} + \rho_{\Delta^0}^{} + \rho_{\Delta^+}^{} +
\rho_{\Delta^{++}}^{}$ and the isospin asymmetry $\delta_{\rm
like}=(\rho_n^{} - \rho_p^{} + \rho_{\Delta^-}^{} -
\rho_{\Delta^{++}}^{}+ \rho_{\Delta^0}^{}/3 -
\rho_{\Delta^+}^{}/3)/\rho_B^{}$ of the asymmetric hadronic matter,
which is a generalization of the isospin asymmetry $\delta =
(\rho_n^{} - \rho_p^{})/(\rho_n^{} + \rho_p^{})$ usually defined for
asymmetric nuclear matter without $\Delta$ resonances~\cite{bali02}.
The nucleon mean-field potential also depends on the stiffness of
nuclear symmetry energy through the parameter $x$ via the functions
$F(x)$ and $G(x)$. We consider the three cases of $x=0$, $x=0.5$,
and $x=1$ with corresponding values $F(x=0) = 129.98$ and $G(x=0) =
1.059$, $F(x=0.5) = 85.54$ and $G(x=0.5) = 1.212$, and $F(x=1) =
107.23$ and $G(x=1) = 1.246$. The resulting nuclear symmetry energy
becomes increasingly softer as the value of $x$ increases, with
$x=1$ giving a nuclear symmetry energy that becomes negative at
about 3 times the normal nuclear matter density. These symmetry
energies reflect the uncertainties in the theoretical predictions on
the stiffness of nuclear symmetry energy at high densities. For the
mean-field potentials of $\Delta$ resonances, their isoscalar
potentials are assumed to be the same as those of nucleons, and
their symmetry potentials are taken to be the average of those for
neutrons and protons with weighting factors depending on the charge
state of $\Delta$ resonance~\cite{art}, i.e., $U_{\rm
asy}^{\Delta^{++}} = U_{\rm asy}^p$, $U_{\rm asy}^{\Delta^+} =
{\textstyle\frac{2}{3}} U_{\rm asy}^p + {\textstyle\frac{1}{3}}
U_{\rm asy}^n$, $U_{\rm asy}^{\Delta^0} = {\textstyle\frac{1}{3}}
U_{\rm asy}^p + {\textstyle\frac{2}{3}} U_{\rm asy}^n$, and $U_{\rm
asy}^{\Delta^-} = U_{\rm asy}^n$.

Including the short-range $\Delta$-hole repulsive interaction via
the Migdal parameter $g^\prime$, which has values $1/3 \le g^\prime
\le 0.6$~\cite{weise75,friedmann81,oset82,xia94,hees05,korpa08},
modifies the pion self-energy to $\Pi^{m_t} =
\Pi_0^{m_t}/(1-g^\prime\Pi_0^{m_t}/k^2)$. The pion spectral function
$S_\pi^{m_t}(\omega,k)$ is then related to the imaginary part of its
in-medium propagator $D^{m_t}(\omega,k) =
1/[\omega^2-k^2-m_\pi^2-\Pi^{m_t}(\omega,k)]$ via
$S_\pi^{m_t}(\omega,k) = -(1/\pi)\,\mbox{Im}\,D^{m_t}(\omega,k)$.

The modification of the pion properties in nuclear medium affects
the decay width and mass distribution of $\Delta$ resonance. For a
$\Delta$ resonance of isospin state $m_T^{}$ and mass $M$ and at
rest in nuclear matter, its decay width is then given by~\cite{ko89}
\begin{eqnarray}\label{gamma}
&&\Gamma_\Delta^{m_T^{}}(M)\approx -2 \sum_{m_\tau,m_t} |\langle {\textstyle\frac{3}{2}} \,
m_T^{} | 1\, m_t\, {\textstyle\frac{1}{2}} \, m_\tau \rangle|^2 \int \frac{d^3{\bf k}}{(2\pi)^3} \left(
\frac{f_\Delta^{}}{m_\pi^{}} \right)^2 F_\pi^2(k)\nonumber\\
&&\times\left[\frac{1}{z_\pi^{-1}e^{(\omega-m_t\mu_Q^{})/T}-1}+1\right]\left[1-\frac{1}{e^{(m_N^{} + k^2/2m_N^{} +U_N^{m_\tau}-\mu_B^{}-2m_\tau\mu_Q^{})/T}+1}\right]\\
&&\times\mbox{Im}\, \left[\frac{k^2}{3}\frac{D^{m_t}(\omega,k)}
{(1-g^\prime\Pi_0^{m_t}(\omega,k)/k^2)^2} +
{g^\prime}^2\frac{\Pi^{m_t}(\omega,k)}{k^2} \right].\nonumber
\end{eqnarray}
In the above, the first term in the last line is due to the decay of
the $\Delta$ resonance to  pion but corrected by the contact
interaction at the $\pi N\Delta$ vertex, while the second term
contains the contribution from its decay to the $\Delta$-hole state
without coupling to  pion. The first two factors in the momentum
integral take into account, respectively, the Bose enhancement for
the pion and the Pauli blocking of the nucleon. To
include possible chemical non-equilibrium effect, a fugacity
parameter $z_\pi^{}$ is introduced for pions.
The pion energy $\omega$ is determined from energy conservation,
i.e., $M + U_\Delta^{m_T^{}} = \omega + m_N^{} + k^2/2m_N^{} +
U_N^{m_\tau}$. The resulting mass distribution of $\Delta$
resonances is then given by $P_\Delta(M) =
A[\Gamma_\Delta^{m_T^{}}(M)/2]/[(M-m_\Delta^{})^2
+{\Gamma_\Delta^{m_T^{}}}^2(M)/4]$, where $A$ is a normalization
constant to ensure the integration of $P_\Delta(M)$ over $M$ is one.

\begin{figure}[h]
\centerline{\includegraphics[width=4.5in,height=4.5in,angle=0]{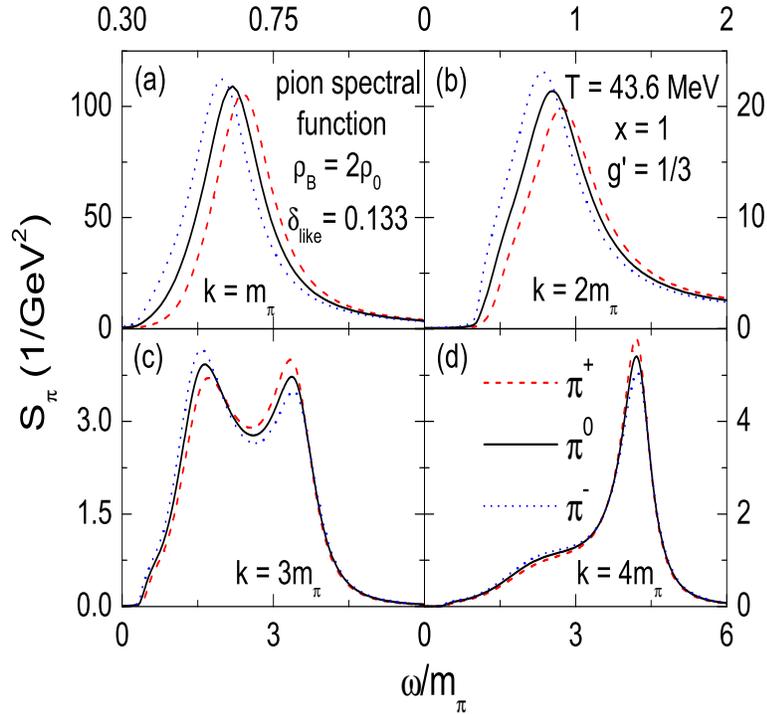}}
\caption{(Color online) Spectral functions of pions in asymmetric
nuclear matter of density $2\rho_0^{}$ and isospin asymmetry
$\delta_{\rm like}=0.133$ as functions of pion energy for different
pion momenta of (a) $m_\pi$, (b) $2 m_\pi$,  (c) $3 m_\pi$, and (d)
$4 m_\pi$. All are calculated with the Migdal parameter
$g^\prime=1/3$.}\label{pion}
\end{figure}

\begin{figure}[h]
\centerline{\includegraphics[width=3.5in,height=3.5in,angle=0]{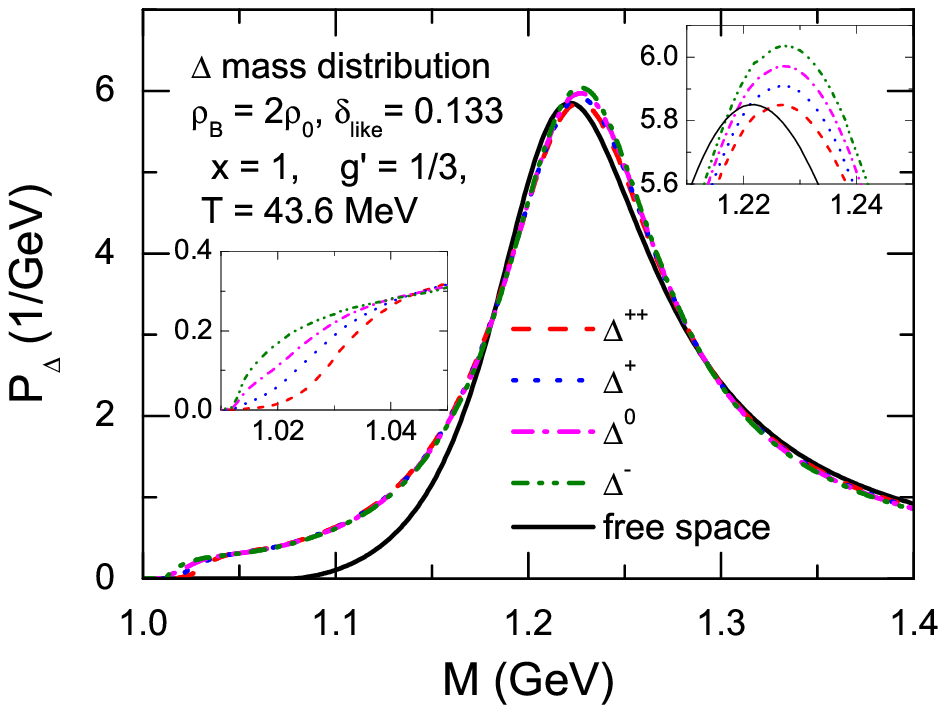}}
\caption{(Color online) Mass distributions of $\Delta$ resonances at
rest in asymmetric nuclear matter of density $2\rho_0^{}$ and
isospin asymmetry $\delta_{\rm like}=0.133$. The solid line
corresponds to that in free space. The distributions near the
threshold and at the peak are enlarged in the insets.} \label{delta}
\end{figure}

We have solved Eqs.~(\ref{pi}) and (\ref{gamma}) self-consistently
to obtain the pion spectral functions and the mass distributions of
$\Delta$ resonances in asymmetric nuclear matter. The results
obtained with the Migdal parameter $g^\prime = 1/3$ are illustrated
in Fig.~\ref{pion} and Fig.~\ref{delta} for an asymmetric nuclear
matter of isospin asymmetry $\delta_{\rm like}\simeq 0.133$, twice
the normal nuclear matter density $\rho_B^{} = 2\rho_0^{}$,
temperature $T\simeq 43.6~{\rm MeV}$, and chemical potentials
$\mu_B^{} \simeq 941.89~{\rm MeV}$ and $\mu_Q^{} \simeq -18.26~{\rm
MeV}$, corresponding to those to be used in our thermal model and
also similar to those reached in the transport model with the
nuclear symmetry energy $x=1$ for central Au+Au collisions at the
beam energy of $0.4~{\rm AGeV}$~\cite{xiao09}. Shown in
Fig.~\ref{pion} are the pion spectral functions as functions of pion
energy for different values of pion momentum. It is seen that for
low pion momenta the spectral function at low energies has a larger
strength for $\pi^-$ (dotted line) than for $\pi^0$ (solid line),
which has a strength larger than that for $\pi^+$ (dashed line).
This behavior is reversed for high pion energies. Fig.~\ref{delta}
shows the mass distributions of $\Delta$ resonances at rest in
asymmetric nuclear matter as functions of mass. One sees that they
are similar to that in free space (solid line) as a result of the
cancelation between the pion in-medium effects, which
enhance the strength at low masses, and the Pauli-blocking
of the nucleon from delta decay, which reduces the strength at low
masses. This is consistent with the observed similar energy
dependence of the photo-proton and photo-nucleus absorption cross
sections around the $\Delta$ resonance mass~\cite{vanPee:2007tw}.
Furthermore, the strength around the peak and near the threshold of
the $\Delta$ resonance mass distribution slightly decreases with
increasing charge of the $\Delta$ resonance due to nonzero isospin
asymmetry of the nuclear medium.

\section{Charged pion ratio in hot dense asymmetric nuclear matter}

To see the above isospin-dependent pion in-medium effects on the
$\pi^-/\pi^+$ ratio in heavy ion collisions, we have used a thermal
model which assumes that pions are in thermal equilibrium with
nucleons and $\Delta$ resonances~\cite{bertsch}. In terms of the
spectral function $S_i(\omega,k)$, the density of a particle species
$i$ is then given by
\begin{eqnarray}\label{density}
\rho_i^{} \approx g_i^{} \int \frac{d^3{\bf k}}{(2\pi)^3}
d\omega^{n_i^{}} S_i(\omega,k)\frac{1}{z_i^{-1}
e^{(\omega - B_i\mu_B^{}-Q_i\mu_Q^{} )/T} \pm 1}.
\end{eqnarray}
In the above, $g_i^{}$, $B_i$, and $Q_i$ are the degeneracy, baryon
number, and charge of the particle. The fugacity parameter
$z_i^{}$ is introduced to take into account possible chemical non-equilibrium effect.
The exponent $n_i^{}$ is $2$ for pions and $1$ for nucleons
and $\Delta$ resonances. For the spectral functions of $\Delta$
resonances, we neglect their momentum dependence and replace the
integration over the energy $\omega$ by that over mass. The $\omega$
in the Fermi-Dirac distribution for $\Delta$ resonances is then
simply $\omega=M+k^2/2M+U_\Delta^{m_T}$. For nucleons, their
spectral functions are taken to be delta functions if we neglect the
imaginary part of their self-energies, i.e., $S_N^{m_\tau}(\omega,k)
= \delta ( \omega^{} - m_N^{} - k^2/2m_N^{} - U_N^{m_\tau} ).$

According to studies based on the transport
model~\cite{bali02,xiao09,xiong93}, the total number of pions and
$\Delta$ resonances in heavy ion collisions reaches a maximum value
when the colliding matter achieves the maximum density, and remains
essentially constant during the expansion of the matter. For Au+Au
collisions at the beam energy of $0.4~{\rm AGeV}$, for which the
$\pi^-/\pi^+$ ratio has been measured by the FOPI Collaboration at
GSI~\cite{FOPI}, the IBUU transport model gives a maximum density
that is about twice the normal nuclear matter density and is
insensitive to the stiffness of the nuclear symmetry energy, as it
is mainly determined by the isoscalar part of the nuclear equation
of state~\cite{xiao09}. This density is thus used in the thermal
model. The temperature in the thermal model is determined by fitting
the measured pion to nucleon ratio, which is about $0.014$ including
pions and nucleons from the decay of $\Delta$
resonances~\cite{FOPI}, without medium effects and with unity
fugacity parameters for all particles, and the value is $T \simeq
43.6$~MeV. The assumption that pions and $\Delta$ resonances are in
chemical equilibrium is consistent with the short chemical
equilibration times estimated from the pion and $\Delta$ resonance
production rates. The isospin asymmetry of the hadronic matter is
then taken to be $\delta_{\rm like} \simeq 0.080$, $0.106$, and
$0.143$, corresponding to net charge densities of $0.920\rho_0^{}$,
$0.894\rho_0^{}$ and $0.857\rho_0^{}$, for the three symmetry
energies given by $x=0$, $0.5$, and $1$, respectively, in order to
reproduce the $\pi^-/\pi^+$ ratios of $2.20$, $2.40$, and $2.60$
predicted by the IBUU transport model of Ref.~\cite{xiao09} using
corresponding symmetry energy parameters without pion in-medium
effects. Since the medium effects enhance the pion and $\Delta$
resonance densities, to maintain the same pion to nucleon ratio as
the measured one requires  the fugacity parameters for pions and
$\Delta$ resonances to be less than one. Also, the pion in-medium
effects have been shown to affect only slightly the pion and the
$\Delta$ resonance abundance~\cite{xiong93}, indicating that both
pions and $\Delta$ resonances are out of chemical equilibrium with
nucleons when medium effects are included, as expected from the
estimated increasing pion and $\Delta$ resonance chemical
equilibration times as a result of the medium effects. Because of
the small number of pions (about 0.3\%) and $\Delta$ resonances
(about 1.1\%) in the matter, the density, temperature, and net
charge density of the hadronic matter are expected to remain
unchanged when the pion in-medium effects are introduced. They lead
to, however, a slight reduction of the isospin asymmetry to
$\delta_{\rm like} \simeq 0.073$, $0.097$, and $0.133$ for the three
symmetry energies, respectively. We note that with the fugacity of nucleons
kept at $z_N=1$, the fugacity parameters of about $z_\pi^{} = 0.061$ and
$z_\Delta^{}= 0.373$ are needed to maintain same total number of pions and $\Delta$ resonances
as in the case without pion in-medium effects, and that the required values
for the fugacity parameters increase only slightly for the other two
symmetry energies considered here.

\begin{figure}[h]
\centerline{\includegraphics[width=3.5in,height=3.5in,angle=0]{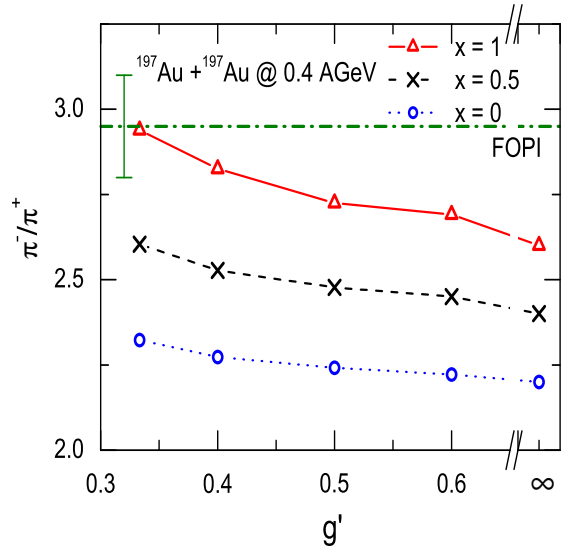}}
\caption{(Color online) The $\pi^-/\pi^+$ ratio in Au+Au collisions
at the beam energy of $0.4~{\rm AGeV}$ for different values of
nuclear symmetry energy ($x=0$, $0.5$, and $1$) and the Migdal
parameter $g^\prime=1/3$, $0.4$, $0.5$, and $0.6$ in the
$\Delta$-hole model for the pion $p$-wave interaction.  Results for
$g^\prime=\infty$ correspond to the case without the pion in-medium
effects.} \label{ratio}
\end{figure}

Results on the $\pi^-/\pi^+$ ratio in Au+Au collisions at the beam
energy of $0.4~{\rm AGeV}$ are shown in Fig.~\ref{ratio}. With the
value $g^\prime = 1/3$ for the Migdal parameter, values for the
$\pi^-/\pi^+$ ratio are $2.32$, $2.60$, and $2.94$ for the
three symmetry energy parameters $x=0$, $0.5$, and $1$,
respectively, which are larger than corresponding values for the
case without including the pion in-medium effects as shown by those
for $g^\prime = \infty$ in Fig.~\ref{ratio}. These results indicate
that the isospin-dependent pion in-medium effects on the charged
pion ratio are comparable to those due to the uncertainties in the
theoretically predicted stiffness of the nuclear symmetry energy.
The measured $\pi^-/\pi^+$ ratio of about $3$ by the FOPI
Collaboration, shown in Fig.~\ref{ratio} by the dash-dotted line
together with the error bar, which without the pion in-medium
effects favors a nuclear symmetry energy softer than the one given
by $x=1$, is now best described by a less softer one.

Fig.~\ref{ratio} further shows the results obtained with larger
values of $g^\prime=0.4$, $0.5$ and $0.6$ for the Migdal parameter.
It is seen that the isospin-dependent pion in-medium effects are
reduced in these cases compared to the case of $g^\prime = 1/3$ as
the repulsive interaction between $\Delta$-hole states becomes
stronger, thus reducing the pion in-medium effects. With these
larger values of $g^\prime$, symmetry energies softer than that
given by $x=1$ are then needed to describe the measured
$\pi^-/\pi^+$ ratio.

\section{Pion $s$-wave interactions in nuclear medium}

The above study does not include the $s$-wave interactions of
pions with nucleons. Calculations based on the chiral perturbation
theory have shown that the pion $s$-wave interaction modifies the
mass of a pion in nuclear medium, and for asymmetric nuclear matter
this effect depends on the charge of the pion~\cite{Kaiser:2001bx}.
Up to the two-loop approximation in chiral perturbation
theory~\cite{Kaiser:2001bx}, the self energies of $\pi^-$, $\pi^+$,
and $\pi^0$ in asymmetric nuclear matter of proton density $\rho_p$
and neutron density $\rho_n$ are given, respectively, by
\begin{eqnarray}
\Pi^-(\rho_p,\rho_n)&=&\rho_n[T^-_{\pi N}-T^+_{\pi N}]-\rho_p[T^-_{\pi N}+T^+_{\pi N}]+\Pi^-_{\rm rel}(\rho_p,\rho_n)+\Pi^-_{\rm cor}(\rho_p,\rho_n)\notag\\
\Pi^+(\rho_p,\rho_n)&=&\Pi^-(\rho_n,\rho_p)\notag\\
\Pi^0(\rho_p,\rho_n)&=&-(\rho_p+\rho_n)T^+_{\pi N}+\Pi^0_{\rm cor}(\rho_p,\rho_n).
\end{eqnarray}
In the above, $T^\pm$ are the isospin-even and isospin-odd $\pi
N$-scattering $T$-matrices which have the empirical values $T^-_{\rm
\pi N}\approx 1.847~{\rm fm}$ and $T^+_{\rm \pi N}\approx
-0.045~{\rm fm}$ extracted from the energy shift and width of the 1s
level in pionic hydrogen atom. The term $\Pi^-_{\rm rel}$ is due to
the relativistic correction, whereas the terms $\Pi^-_{\rm cor}$ and
$\Pi^0_{\rm cor}$ are the contributions from the two-loop order in
chiral perturbation theory.  Numerically, it was found in
Ref.~\cite{Kaiser:2001bx} that changes of pion masses in asymmetric
nuclear matter of density $\rho=0.165~{\rm fm}^{-3}$ and isospin
asymmetry $\delta=0.2$ are $\Delta m_{\pi^-}=13.8~{\rm MeV}$,
$\Delta m_{\pi^+}=-1.2~{\rm MeV}$, and $\Delta
m_{\pi^0}=6.1~{\rm MeV}$.

\begin{figure}[h]
\centerline{\includegraphics[width=3.5in,height=3.5in,angle=0]{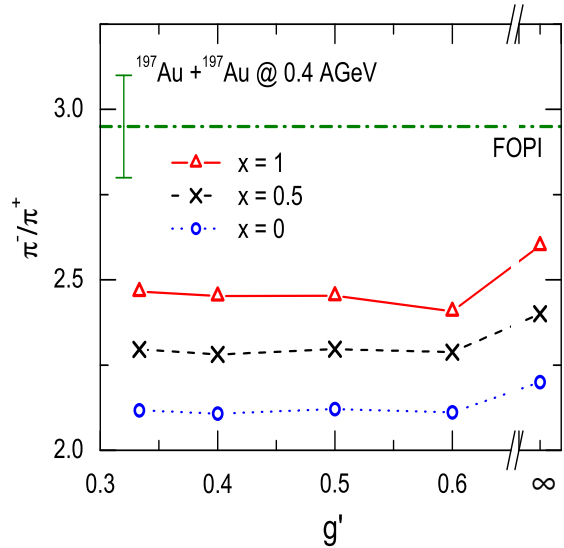}}
\caption{(Color online) Similar to Fig.~\ref{ratio} with both pion
$s$-wave and $p$-wave interactions included.}\label{ratiosp}
\end{figure}

Taking into account the isospin-dependent pion self energies due to
pion $s$-wave interactions in asymmetric nuclear matter changes the
results shown in Figs.~\ref{pion} and \ref{delta}. For the pion
spectral function, the one for $\pi^+$ now has a larger strength at
low energies than that for $\pi^-$. Similarly, the strength near the
threshold of the $\Delta$ resonance mass distribution now increases
with increasing charge of the $\Delta$ resonance, although that around
the peak still decreases with increasing $\Delta$ resonance charge.  As a result,  the
$\pi^-/\pi^+$ ratio in Au+Au collisions at the beam energy of
$0.4~{\rm AGeV}$ is slightly reduced after the inclusion of both
pion $s$-wave and $p$-wave interactions in asymmetric nuclear matter
as shown in Fig.~\ref{ratiosp}.

\section{Summary}

The pion spectral function in asymmetric nuclear matter becomes
dependent on the charge of a pion. For the $p$-wave interaction of
the pion, modeled by its couplings to the $\Delta$-hole excitations
in nuclear medium, it leads to an increased strength of the $\pi^-$
spectral function at low energies relative to that of the $\pi^+$
spectral function in dense asymmetric nuclear matter.  In a thermal
model, this isospin-dependent effect increases the $\pi^-/\pi^+$
ratio from heavy ion collisions, and the effect is comparable to
that due to the uncertainties in the theoretically predicted
stiffness of the nuclear symmetry energy at high densities. However,
including also the pion $s$-wave interaction based on results from
the chiral perturbation theory reverses the isospin-dependent pion
in-medium effects, leading instead to a slightly reduced
$\pi^-/\pi^+$ ratio in neutron-rich nuclear matter. Taking
into consideration of the isospin-dependent pion in-medium effects
in the transport model thus would have some influence on the extraction of the
nuclear symmetry energy from measured $\pi^-/\pi^+$ ratio.

\section*{Acknowledgments}

This talk was based on work supported in part by the US National Science
Foundation under Grant No. PHY-0758115 and the Welch Foundation
under Grant No. A-1358.


\begin{thebibliography}{10}

\bibitem{LCK08}
B.~A. Li, L.~W. Chen, and C.~M. Ko, {\it Phys. Rep.} \textbf{464} (2008) 113.

\bibitem{ireview98}
B.~A. Li, C.~M. Ko, and W.~Bauer, {\it Int. Jour. Mod. Phys. E} {\bf 7} (1998) 147.

\bibitem{ibook}
{\it Isospin Physics in Heavy-Ion Collisions at Intermediate Energies},
eds. Bao-An Li and W. Udo Schr\"{o}der (Nova Science Publishers, Inc, New York, 2001).

\bibitem{baran05}
V.~Baran, M.~Colonna, V.~Greco, and M. Di~Toro, {\it Phys. Rep.} \textbf{410} (2005) 335.

\bibitem{Tsa04}
M.~B. Tsang {\it et al.\/}, {\it Phys. Rev. Lett.} \textbf{92} (2004) 062701.

\bibitem{Liu07}
T.~X. Liu {\it et al.\/}, {\it Phys. Rev. C} \textbf{76} (2007)
034603\com{.}

\bibitem{Che05a}
L.~W. Chen, C.~M. Ko, and B.~A. Li, {\it Phys. Rev. Lett.} \textbf{94} (2005) 032701.

\bibitem{LiBA05c}
B.~A. Li and L.~W. Chen, {\it Phys. Rev. C} {\bf 72} (2005) 064611.

\bibitem{She07}
D.~V. Shetty, S.~J. Yennello, and G.~A. Souliotis, {\it Phys. Rev.
C} {\bf 75} (2007) 034602.

\bibitem{Ste05b}
A.~W. Steiner and B.~A. Li, {\it Phys. Rev. C} {\bf 72} (2005) 041601(R).

\bibitem{Gar07}
T.~Li {\it et al.\/}, {\it Phys. Rev. Lett.} {\bf 99} (2007) 162503.

\bibitem{bali02}
B.~A. Li, {\it Phys. Rev. Lett.} {\bf 88} (2002) 192701;
{\it Nucl. Phys. A} {\bf 708} (2002) 365.

\bibitem{Ferini:2006je}
G.~Ferini {\it et al.}, {\it Phys.\ Rev.\ Lett.}  {\bf 97} (2006) 202301.

\bibitem{xiao09}
Z.~G. Xiao, B.~A. Li, L.~W. Chen, G.~C. Yong, and M.~Zhang,
{\it Phys. Rev. Lett.} {\bf 102} (2009) 062502;
M. Zhang, Z.~G. Xiao, B.~A. Li, L.~W. Chen, G.~C. Yong, and S.~J.
Zhu, {\it Phys. Rev. C} {\bf 80} (2009) 034616.

\bibitem{FOPI}
W.~Reisdorf {\it et al.\/} (FOPI Collaboration), {\it Nucl. Phys. A} \textbf{781} (2007) 459.

\bibitem{Ferrini:2005jw}
G.~Ferini, M.~Colonna, T.~Gaitanos and M.~Di Toro,
{\it Nucl.\ Phys.\ A}  {\bf 762} (2005) 147.

\bibitem{weise75}
G.~E. Brown and W.~Weise, {\it Phys. Rep.}  {\bf 22} (1975) 279.

\bibitem{friedmann81}
B.~Friedmann, V.~R. Pandharipandi, and Q.~N. Usmani, {\it Nucl.
Phys. A} {\bf 372} (1981) 483.

\bibitem{oset82}
E.~Oset, H.~Toki, and W.~Weise, {\it Phys. Rep.} {\bf 83} (1982) 281.

\bibitem{xia94}
L.~H. Xia, P.~Siemens, and M.~Soyeur, {\it Nucl. Phys. A} {\bf 578} (1994) 493.

\bibitem{hees05}
H.~Van Hees and R.~Rapp, {\it Phys. Lett. B} {\bf 606} (2005) 59.

\bibitem{korpa08}
C.~L. Korpa, M.~F.~M. Lutz, and F.~Riek, {\it Phys. Rev. C} {\bf 80} (2009) 024901.

\bibitem{xiong93}
L.~Xiong, C.~M. Ko, and V.~Koch, {\it Phys. Rev. C} {\bf 47} (1993) 788.

\bibitem{korpa99}
C.~L. Korpa and A.~E.~L. Dieperink, {\it Phys. Lett. B} {\bf 446} (1999) 15.

\bibitem{ko89}
C.~M. Ko, L.~H. Xia, and P.~J. Siemens, {\it Phys. Lett. B} {\bf 231} (1989) 16.

\bibitem{art}
B.~A. Li and C.~M. Ko, {\it Phys. Rev. C}  {\bf 52} (1995) 2037;
B.~A. Li, A.~T. Sustich, B.~Zhang, and C.~M. Ko, {\it Int. Jour. Mod. Phys. E} {\bf 10} (2001) 267.

\bibitem{vanPee:2007tw}
H.~van Pee {\it et al.}  (CB-ELSA Collaboration),
{\it Eur.\ Phys.\ J.\  A} {\bf 31} (2007) 61.

\bibitem{bertsch}
G.~F. Bertsch, {\it Nature} {\bf 283} (1980) 280;
A.~Bonasera and G.~F. Bertsch, {\it Phys. Lett. B} {\bf 195} (1987) 521.

\bibitem{Kaiser:2001bx}
N.~Kaiser and W.~Weise,
{\it Phys.\ Lett.\  B} {\bf 512} (2001) 283.

\end{thebibliography}
\end{document}